\documentclass{jfm}

\usepackage{graphicx}
\usepackage{newtxtext}
\usepackage{newtxmath}
\usepackage{natbib}
\usepackage{hyperref}
\usepackage{soul} 
\usepackage{color}
\usepackage{enumitem}
\usepackage{setspace}
\hypersetup{
    colorlinks = true,
    urlcolor   = blue,
    citecolor  = black,
}

\newcommand{\RomanNumeralCaps}[1]

\title{Distributed force element decomposition with method of fundamental solutions
}

\author{Zhiteng Zhou\aff{1,2}, Yi Liu\aff{1,2}, Hongping Wang\aff{1,2}, \and Shizhao Wang\aff{1,2}\corresp{\email{wangsz@lnm.imech.ac.cn}},
 }

\affiliation{\aff{1}State Key Laboratory of Nonlinear Mechanics, Institute of Mechanics, Chinese Academy of Sciences, Beijing, 100190 , China
\aff{2}School of Engineering Sciences, University of Chinese Academy of Sciences, Beijing, 100049, China}

\begin{document}
\maketitle

\begin{abstract}
Quantifying the contribution of vortex structures to pressure stress is useful for designing flow control strategies to mitigate low or drag. The traditional force-element method focuses on the contribution of vortex structures to the resultant force. However, the contribution of vortex structures to the distributed force can not be identified. To address this problem, a distributed force element method is proposed. This method projects the Navier-Stokes equation to a divergence-free space and obtain a Poisson equation for pressure. The Green's function is employed to express pressure stress on the solid boundary as a combination of volume and surface integrals. Contributions to the distributed source are divided into acceleration surface elements, vorticity surface elements, convection volume elements. The method of fundamental solutions and singular value decomposition are used to efficiently solve the Green's functions.
The distributed force element method is validated using laminar flows around a stationary circular cylinder and oscillating circular cylinder, and three-dimensional laminar and turbulent flows around a sphere. In laminar flow over a circular cylinder, fluctuation of forces mainly originate from volume sources, while in flow over an oscillating  circular cylinder, inertial and volume source terms may cancel out, suppressing lift fluctuation. In flows over a sphere, a strong negative volume source in the boundary layer on the sphere creates low-pressure regions, while pressure in the separated shear layer is close to zero.

\end{abstract}


\section{Introduction}

Quantitatively relating vortex structures to forces is valuable for gaining insight into the dominant mechanisms governing the resistance and noise~\citep{wu2018fundamental}. For instance, the relationship between the vortex structure and resultant force reveals that the force on a flapping plate is determined solely by the two attached wake vortices~\citep{li2012force}. The analysis of lamb vector in flow over bluff bodies assists to reduction of bluff bodies'
resistance  
~\citep{chang2008many}.
The revelation of the quantitative relationship between pressure gradients and vorticity generation aids in the control of flow separation~\citep{zhu2015causal}.


Different methods have been developed to quantify the relationship between vortex structures and resultant forces. One approach uses impulse theory to link bulk flow features to resultant force, which is called vorticity-moment method~\citep{wu2018fundamental}. The vorticity-moment theory traces its origins to foundational work in the early 20th century. \cite{WOS:000202536000073} pioneered the concept by linking aerodynamic forces to the time rate of change of fluid impulse, proposing that vorticity generation and diffusion at the body surface govern resistance. This idea remained dormant until~\cite{wu1981theory} and~\cite{lighthill1986informal} revitalized and systematized the theory, establishing its modern framework for incompressible flows. They demonstrated that the resultant force on a body in unbounded flow could be expressed through a global integral of vorticity moments, bypassing direct pressure calculations.
By using derivative-moment transformations, Wu et al.~\cite{wu2005unsteady} proposed a formula for calculating the resultant force that relies exclusively on the boundary integrals over a control surface enclosing the solid body.
The theory gained traction in biological propulsion studies, where discrete vortex structures dominate unsteady flows.
~\cite{li2012force} proposed a finite-domain impulse formulation and found that the resultant force on a flapping plate is determined solely by the two wake vortices that remain connected to the body.
~\cite{kang2018minimum} later improved the theory by developing a minimum-domain approach, showing that forces could be determined using only the vorticity attached to or near the body, dramatically simplifying applications to complex wake interactions. Recent applications by~\cite{liu2015lift} bridged classical linear models with viscous flows, revealing how finite Reynolds number effects alter vortex shedding patterns. 

Another approach relies on projecting the Navier-Stokes (NS) equation to an irrotational space to decompose the flow fields into different components, known as force-element theory. The force-element theory decouples pressure from vortex structures~\citep{quartapelle1983force}, enabling direct calculation of resultant force via vorticity~\citep{chang1992potential,chang2008many}. 
The force-element theory originated in the early 1980s  frrom the work of~\cite{quartapelle1983force}, who introduced auxiliary divergence-free vector fields to reformulate the pressure integral into volume integrals of acceleration and vorticity terms. ~\cite{chang1992potential} further refined the theory by decomposing forces into inertial (added-mass), vortical (Lamb vector), and viscous components, enabling clearer physical interpretation. ~\cite{howe1995force} linked the theory to vortex-sound interactions, using potential-based basis functions tied to body geometry to unify force and sound generation mechanisms. ~\cite{hsieh2009investigation} applied the theory to biological locomotion, demonstrating its capability in analyzing unsteady flows by isolating dominating force contributors. 
Traditional potential-based methods, relying on Gauss's theorem, primarily resolve resultant force  rather than local variations, which inherently limits their ability to connect vortex structures with distributed forces. 

This work proposes a distributed force element method based on projection of NS equation to a divergence-free space. Unlike conventional approaches, the proposed method decomposes pressure stress into discrete ``elements'' tied to specific volume and surface sources. The Green's function method is introduced to represent the contribution weights of sources at different locations to pressure stress. To reduce the computational cost required by the Green's function, this work employs the method of fundamental solutions to decrease the size of the equation system to be solved, and further transforms the matrix solution into vector multiplication by using singular value decomposition.

The work is organized as follows: The details of the proposed distributed force element method are given in Section 2. Validations of the proposed approach with 2D and 3D flows over a static or moving solid body are presented in Section 3. Finally, the conclusions are drawn in Section 4. 

\section{ Pressure stress in flows}
According to Navier-Stokes equations for compressible flows, the governing equation of pressure $p$ can be written as
\begin{equation}\label{eqk0}
\begin{split}
\frac{1}{{c_0^2}}\frac{{{\partial ^2}p}}{{\partial {t^2}}} - \frac{{{\partial ^2}p}}{{\partial {x_i}\partial {x_i}}} = \frac{{{\partial ^2}\left( {\rho {u_i}{u_j} + \left( {\left( {p - {p_0}} \right) - {c^2}\left( {\rho  - {\rho _0}} \right)} \right){\delta _{ij}} - {\tau _{ij}}} \right)}}{{\partial {x_i}\partial {x_j}}}
\end{split}
\end{equation}
under homentropic assumption~\citep{lighthill1952sound,glegg2017aeroacoustics}. $\rho$ denotes density. $c_0$, $p_0$ and $\rho_0$ are the speed of sound, pressure and density in the freestream flow. 
${u_i}$ and ${u_j}$ are the i-th and j-th components of fluid velocity. 
${\delta _{ij}}$ represents the Kronecker delta  function and ${\tau _{ij}}$ the viscous stress. 

\cite{kambe1993oblique} divided the pressure into inner and outer regions in low Mach number and high Reynolds number flows according to the distance from the source. In the outer region away from the source, the pressure is governed by the acoustic wave propagation. The source in the right-hand-side of Eq.~(\ref{eqk0}) is negligible and we have
\begin{equation}\label{eqk1}
\begin{split}
\frac{1}{{c_0^2}}\frac{{{\partial ^2}p}}{{\partial {t^2}}} - \frac{{{\partial ^2}p}}{{\partial {x_i}\partial {x_i}}} = 0
\end{split}
\end{equation}
when \( \text{Re} \gg 1 \).
In the inner region near the source, the pressure is governed by vortex dynamics. Since 
$
\frac{\partial^2 p}{c^2 \partial t^2} / \Delta p = O(M^2),
$ where $M$ is the freestream Mach number, $\frac{1}{{c_0^2}}\frac{{{\partial ^2}p}}{{\partial {t^2}}}$ can be neglected in the left-hand-side of Eq.~(\ref{eqk0}). Meanwhile
the entropy fluctuation 
$
(p - p_0) - c^2 (\rho - \rho_0)
$
can be neglected at low Mach numbers, and the viscous stress 
$
\tau_{ij}
$
can be neglected when \( \text{Re} \gg 1 \). Therefore, Eq.~(\ref{eqk0}) can be reduced to
\begin{equation}\label{eqk2}
\begin{split}
\Delta p =  - \frac{{{\partial ^2}\left( {\rho {u_i}{u_j}} \right)}}{{\partial {x_i}\partial {x_j}}}
\end{split}
\end{equation}
in low Mach number turbulence. The Eqs.~(\ref{eqk1}) and~(\ref{eqk2}) can be solved by matching the expanded pressure in the intermediate region where the distance to the source is much greater than the vortex scale and much less than the sound wave length.
The coefficients of the pressure's asymptotic expansion in the inner region are determined by the pressure in the outer region.

%
The aforementioned method has been employed in the study of aerodynamic sound~\citep{kambe1993oblique,zang2025}. In this work, the method is applied to the decomposition of pressure stress on the solid boundary within the inner region.
The Neumann condition at the solid boundary is required to specify the pressure uniquely. By projecting the Navier-Stokes equations to the norm ${\bf{n}}$ of the solid boundary, the Neumann condition can be obtained as
\begin{equation}\label{eq2}
\begin{split}
\frac{{\partial p}}{{\partial n}} = \left( { - \frac{1}{{{\mathop{\rm Re}\nolimits} }}\nabla  \times {\bf{\omega }}} \right) \cdot {\bf{n}} - \frac{{D{\bf{u}}}}{{Dt}}\cdot{\bf{n}} ,
\end{split}
\end{equation}
where the first term on the right-hand-side ${ \frac{-\nabla  \times {\bf{\omega }}}{{{\mathop{\rm Re}\nolimits} }}}$ denotes the effects of vorticity and the second term $\frac{{-D{\bf{u}}}}{{Dt}}$ denotes the effects of  acceleration of the solid boundary.  


\section{Pressure decomposition}
According to the boundary condition and Poisson equation for the pressure, a decomposition on the pressure can be proposed as 
\begin{equation}\label{eq2-1}
\begin{split}
p = p_f + p_s,
\end{split}
\end{equation} 
where $p_f$ is the pressure from volume sources and satisfies
\begin{equation}\label{eq3}
\begin{split}
\nabla^2 p_f=- \nabla \cdot(\boldsymbol{{\bf{u}}} \cdot \nabla \boldsymbol{{\bf{u}}}),
\end{split}
\end{equation}
in the fluid and 
\begin{equation}\label{eq4}
\begin{split}
\frac{{\partial p_f}}{{\partial n}} = 0,
\end{split}
\end{equation}
on the solid boundary, while $p_s$ is the pressure from surface sources and satisfies
\begin{equation}\label{eq5}
\begin{split}
\nabla^2 p_s=0,
\end{split}
\end{equation}
in the fluid and 
\begin{equation}\label{eq6}
\begin{split}
\frac{{\partial p_s}}{{\partial n}} =  \left( { - \frac{1}{{{\mathop{\rm Re}\nolimits} }}\nabla  \times {\bf{\omega }}} \right) \cdot {\bf{n}} - \frac{{D{\bf{u}}}}{{Dt}} \cdot {\bf{n}},
\end{split}
\end{equation}
on the solid boundary.
Inspired by the work of~\cite{curle1955influence} and~\cite{chang1992potential,chang2008many}, we define Green's function $G_f$ as ${\nabla ^2}{G_f} = \delta ({\bf{x}} - {\bf{y}}),{\rm{ }}\partial {G_f}/\partial n = 0$ and $G_s$ as ${\nabla ^2}{G_s} = 0,{\rm{ }}\partial {G_s}/\partial n = \delta ({\bf{x_s}} - {\bf{y_s}})$.A equation to obtain $p_f$ can be formulated by multiplying Eq.~(\ref{eq3})  with the corresponding Green's function $G_f$ and subtracting it from the $p_f$ times the $G_f$'s Laplacian.
\begin{equation}\label{eq6-1}
\begin{split}
{p_f}\frac{{{\partial ^2}{G_f}}}{{\partial {y_i}\partial {y_i}}} - {G_f}\frac{{{\partial ^2}{p_f}}}{{\partial {y_i}\partial {y_i}}} = {p_f}\delta ({\bf{x}} - {\bf{y}}) - {G_f}\left( { - \left( {\nabla  \cdot ({\bf{u}} \cdot \nabla {\bf{u}})} \right)} \right).
\end{split}
\end{equation}
Integrating Eq.~(\ref{eq6-1}) of both sides over space yields
\begin{equation}\label{eq6-2}
\begin{split}
&\int_{{V_R}} {\frac{\partial }{{\partial {y_i}}}\left( {{G_f}\frac{{\partial {p_f}}}{{\partial {y_i}}}} \right) - \frac{\partial }{{\partial {y_i}}}\left( {{p_f}\frac{{\partial {G_f}}}{{\partial {y_i}}}} \right)} dV =\\
&\int_{{V_R}} {{G_f}\left( { - \left( {\nabla  \cdot ({\bf{u}} \cdot \nabla {\bf{u}})} \right)} \right)} dV - \int_{{V_R}} {{p_f}\delta ({\bf{x}} - {\bf{y}})} dV.
\end{split}
\end{equation}
Transforming the volume integral in the Eq.~(\ref{eq6-2}) into a surface integral leads to
\begin{equation}\label{eq6-3}
\begin{split}
\int_{{V_R}} {{p_f}\frac{{\partial {G_f}}}{{\partial {y_i}}}{n_i} - {G_f}\frac{{\partial {p_f}}}{{\partial {y_i}}}{n_i}} dS = \int_{{V_R}} {{G_f}\left( { - \left( {\nabla  \cdot ({\bf{u}} \cdot \nabla {\bf{u}})} \right)} \right)} dV - {p_f}.
\end{split}
\end{equation}
Substitute the boundary conditions of \( p_f \) and \( G_f \) into the Eq.~(\ref{eq6-3})  and utilize the sifting property of the \(\delta\) function, we obtain
\begin{equation}\label{eq6-4}
\begin{split}
{p_f} = \underbrace {\int_{{V_R}} { - \left( {\nabla  \cdot ({\bf{u}} \cdot \nabla {\bf{u}})} \right)} {G_f}dV}_{{\rm{convection }}}.
\end{split}
\end{equation}
Similarly, we can follow the steps from  Eq.~(\ref{eq6-1})  to Eq.~(\ref{eq6-3})  to obtain the formula for \( p_s \).
\begin{equation}\label{eq6-5}
\begin{split}
\int_{{V_R}} {{p_s}\frac{{\partial {G_s}}}{{\partial {y_i}}}{n_i} - {G_s}\frac{{\partial {p_s}}}{{\partial {y_i}}}{n_i}} dS = 0.
\end{split}
\end{equation}
Substituting the boundary conditions of \( p_s \) and \( G_s \) into the above equation and utilizing the sifting property of the \(\delta\) function, we obtain
\begin{equation}\label{eq6-6}
\begin{split}
{p_s} = \underbrace { - \int_S {\frac{1}{{{\rm{Re}}}}\left( {\nabla  \times {\bf{\omega }}} \right)}  \cdot {\bf{n}}{G_s}dS}_{{\rm{boundary\; voticity}}}\underbrace { - \int_S {\left( {\frac{{D{\bf{u}}}}{{Dt}} \cdot {\bf{n}}} \right)} {G_s}dS}_{{\rm{acceleration\; of\; solid\; boundary}}}
\end{split}
\end{equation}
on the solid boundary.
The pressure on the solid boundary can thus be obtained by summing Eq.~(\ref{eq6-4}) and Eq.~(\ref{eq6-6}) as
\begin{equation}\label{eq8-1}
\begin{split}
p = \underbrace {\int_{{V_R}} { - \left( {\nabla  \cdot ({\bf{u}} \cdot \nabla {\bf{u}})} \right)} {G_f}dV}_{{\rm{convection }}}\underbrace { - \int_S {\frac{1}{{{\rm{Re}}}}\left( {\nabla  \times {\bf{\omega }}} \right)}  \cdot {\bf{n}}{G_s}dS}_{{\rm{boundary\; voticity}}}\underbrace { - \int_S {\left( {\frac{{D{\bf{u}}}}{{Dt}} \cdot {\bf{n}}} \right)} {G_s}dS}_{{\rm{acceleration\; of\; solid\; boundary}}}.
\end{split}
\end{equation}
Notably in Eq.~(\ref{eq8-1}), the pressure on the solid boundary originates from convection of fluid elements and boundary vorticity and acceleration of the solid boundary.
For convenience of expression, we refer to the source term of \( p_f \) in the flow field as the convection volume source and the source term of \( p_s \) on the solid boundary as the vorticity-acceleration surface source.

\section{Method of fundamental solutions}
To efficiently compute the Green's functions $G_f$ and $G_s$, we introduce the method of fundamental solutions (MFS)~\citep{karageorghis1987method,poullikkas1998methods} and singular value decomposition (SVD). 
The MFS employs {virtual sources} to represent the physical field, where the intensities of these virtual sources are determined by matching the prescribed wall boundary conditions at the collocation points. By using the MFS, the pressure field can be expressed as
\begin{equation}\label{eq9}
\begin{split}
p\left( {\bf{x}} \right) = \sum\limits_j {S({{\bf{y}}_j})} {G}({\bf{x}},{{\bf{y}}_j}),
\end{split}
\end{equation}
where $\mathbf{x},\mathbf{y}$ are the position vectors of observer in the fluid and virtual source inside the solid boundary, respectively. ${{S}({{\bf{y}}_j} )}$ is the strength of the j-th virtual source. $G$ is Green’s function of the Poisson equation in free space. The free-space Green's function naturally meets the far-field conditions.

For a volume source $\delta(\bf{y_f}-\bf{y_0})$ in the fluid, the intensities of the virtual sources can be determined as follows by matching the norm pressure gradient on the solid boundary  (Eq.~(\ref{eq4}))
\begin{equation}\label{eq10}
\begin{split}
\sum\limits_j {S({{\bf{y}}_j},{{\bf{y}}_0})\frac{{\partial G({\bf{x}},{{\bf{y}}_j})}}{{\partial {x_i}}}} {n_i} =  - \frac{{\partial G({\bf{x}},{{\bf{y}}_0})}}{{\partial {x_i}}}{n_i}.
\end{split}
\end{equation}
Then, the Green's function $G_f$ can be obtained as
\begin{equation}\label{eq11}
\begin{split}
{G_f}({\bf{x}},{{\bf{y}}_0}) = G({\bf{x}},{{\bf{y}}_0}) + \sum\limits_j {S({{\bf{y}}_j},{{\bf{y}}_0})} G({\bf{x}},{{\bf{y}}_j}).
\end{split}
\end{equation}
The first term on the right-hand-side of Green's function $G_f$ represents the direct radiation of a volume source on the solid boundary while the second term represents the scatter of the solid boundary. For the convenience of expression, the second term $\sum\limits_j {S({{\bf{y}}_j},\bf{y})} {G}({\bf{x}},{{\bf{y}}_j})$ is defined as  $G_1$. The Green's function $G_f$ is thus  $G_f=G+G_1$. Accordingly, the pressure stress can be further decomposed as follows
\begin{equation}\label{eq12}
\begin{split}
p =& \underbrace {\int_{{V_R}} { - \left( {\nabla  \cdot ({\bf{u}} \cdot \nabla {\bf{u}})} \right)} GdV}_{{\rm{direct}}\;{\rm{radiation}}\;{\rm{of}}\;{\rm{convection}}} + \underbrace {\int_{{V_R}} { - \left( {\nabla  \cdot ({\bf{u}} \cdot \nabla {\bf{u}})} \right)} {G_1}dV}_{{\rm{boundary}}\;{\rm{scatter}}\;{\rm{of}}\;{\rm{convection}}}\\
&\underbrace { - \int_S {\frac{1}{{{\rm{Re}}}}\left( {\nabla  \times {\bf{\omega }}} \right)}  \cdot {\bf{n}}{G_s}dS}_{{\rm{boundary}}\;{\rm{voticity}}}\underbrace { - \int_S {\left( {\frac{{D{\bf{u}}}}{{Dt}} \cdot {\bf{n}}} \right)} {G_s}dS}_{{\rm{acceleration}}\;{\rm{of}}\;{\rm{solid}}\;{\rm{boundary}}}.
\end{split}
\end{equation}
The first and second term in the first line on the right-hand-side of Eq.~(\ref{eq12}) are recorded as $p_{fd}$ and $p_{fs}$ hereinafter.
For a boundary source $\delta(\bf{y_s}-\bf{y_0})$ on the solid boundary, the  intensities of the virtual sources can be determined as follows by using Eq.~(\ref{eq6})
\begin{equation}\label{eq12}
\begin{split}
\sum\limits_j {S({{\bf{y}}_j},{{\bf{y}}_0})\frac{{\partial G({\bf{x}},{{\bf{y}}_j})}}{{\partial {x_i}}}} {n_i} =  \delta(\bf{y_s}-\bf{y_0}).
\end{split}
\end{equation}
Then, the Green's function $G_s$ can be obtained as
\begin{equation}\label{eq13}
\begin{split}
G_s({\bf{x}},{{\bf{y_0}}}) = \sum\limits_j {S({{\bf{y}}_j},\bf{y_0})} {G}({\bf{x}},{{\bf{y}}_j}).
\end{split}
\end{equation}

Eq.~(\ref{eq10}) and ~(\ref{eq12}) can be written in the matrix form as follows
\begin{equation}\label{eq14}
\mathbf{b} = \mathbf{Aq},
\end{equation}
where the matrix $\mathbf{A}$ has elements $A_{mn} = \frac{\partial G(\mathbf{x}_n, \mathbf{y}_m)}{\partial x_i}$ at the $m$-th row and $n$-th column. $\mathbf{q}$ is a column vector, representing the unknown strengths for the virtual sources.
The n-th element in $\mathbf{b}$ is $ - \frac{{\partial G({{\bf{x}}_n},{{\bf{y}}_0})}}{{\partial {x_i}}}{n_i}$ for the volume source and $\delta(\bf{y_s}-\bf{y_0})$ for the surface source. 	
The matrix $\mathbf{A}$ might be rectangular matrices since fewer virtual sources may be used than the collocation points. Therefore, the SVD can be employed to obtain the least-squares solution of the  Eq.~(\ref{eq14}). By using the SVD, the solution of Eq.~(\ref{eq14}) is transformed into vector multiplication and addition as follows
\begin{equation}\label{eq14-1}
\begin{split}
{\bf{q}} \approx \sum\limits_{i = 1}^{{n_c}} {\frac{1}{{{s_i}}}{{\bf{v}}_i}{\bf{u}}_i^Tb}, 
\end{split}
\end{equation}
where $n_c$ is the number of selected modes, taken as the number of virtual sources in this work. $s_i$ denotes the $i$-th singular value of matrix $\mathbf{A}$, while $\mathbf{v}_i$ and $\mathbf{u}_i$ represent the $i$-th right and left singular vectors, respectively.
\section{Results and Discussion}

In this section, the pressure stress computed by using the proposed distributed force element method is compared with the results from directly solving the NS equations. Four cases are investigated below to demonstrate the applicability of the distributed force element method across both two-dimensional (2D) and three-dimensional (3D) flows with stationary or moving boundaries.

\subsection{Two-dimensional flows over a circular cylinder}
 
\begin{figure}
	\centering    \includegraphics[width=0.8\textwidth]{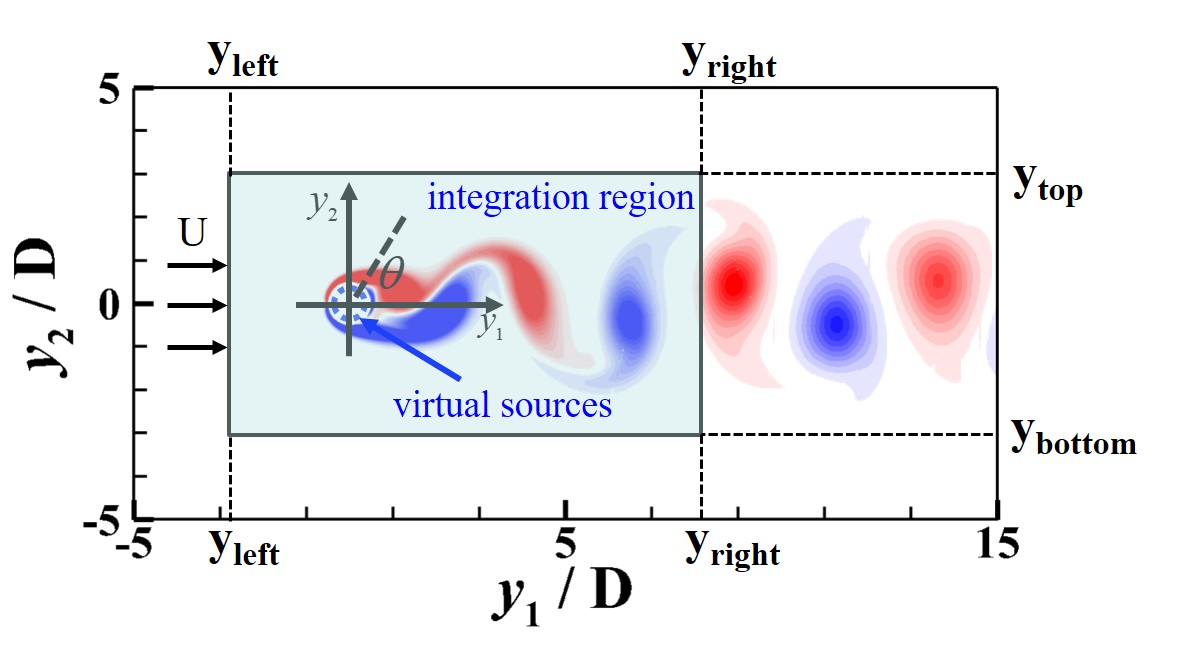}
	\caption{Schematics of the laminar flows over a circular cylinder and setup of the distributed force element method}
	\label{fig1}
\end{figure}

The distributed force element method is first  validated using a two-dimensional flow over a circular cylinder fixed in a uniform freestream flow. As shown in Fig.~\ref{fig1}, the diameter of the circular cylinder is \(D\) and the freestream speed is \(U\). The Reynolds number based on the freestream speed and the diameter of the cylinder is ${\mathop{\rm Re}\nolimits}  = UD/\nu  = 150$, where \(\nu\) is the kinematic viscosity. We set 200 collocation points on the solid boundary and 67 virtual sources, which are uniformly distributed on a circle around the the circular cylinder center. The virtual sources are positioned at $0.3D$ from the circular cylinder center. The integration domain of the volume source is defined as a rectangular region enclosing the cylinder, expressed in coordinates as $[y_{\rm left}, y_{\rm right}] \times [y_{\rm bottom}, y_{\rm top}]$. In this case, $y_{\rm left}$, $y_{\rm bottom}$ and $y_{\rm top}$ are specified as $-2D$, $-5D$ and $5D$, respectively.

The numerical simulation of flow around a circular cylinder is conducted by using a in-house solver for Navier-Stokes equations. This solver employs a second-order Roe scheme for discretizing convective fluxes and a reconstructed central scheme for computing viscous fluxes. An implicit dual-time stepping method of second-order accuracy is utilized for time integration. Additionally, adaptive techniques such as local time-step adjustments and flux blending are incorporated to enhance the stability and convergence of the solution. Comprehensive details regarding the implementation and verification of this numerical approach have been previously documented in our studies~\citep{liu2021numerical,liu2023cache}.

\begin{figure}
	\centering    \includegraphics[width=1\textwidth]{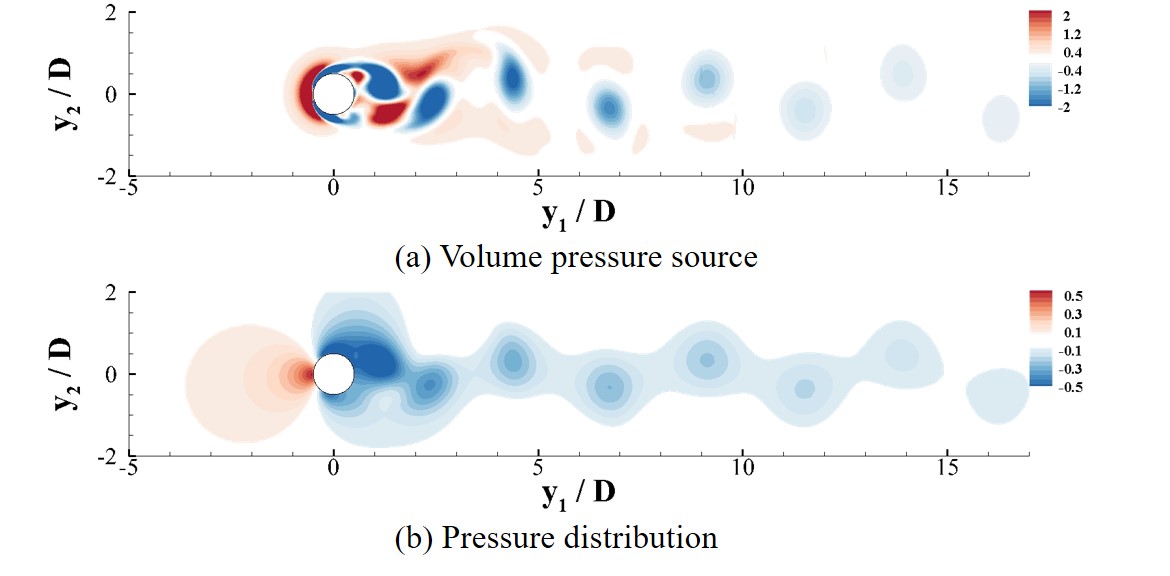}
	\caption{Comparison of the (a) volume source distribution, (b) pressure distribution}
	\label{fig3}
\end{figure}

Figure~\ref{fig3} shows the instantaneous convection volume sources and pressure distribution, corresponding to Figs.~\ref{fig3}a and \ref{fig3}b, respectively. As can be seen from the figures, near the circular cylinder, there is a positive source at the stagnation point in front of the cylinder. Negative volume sources exist on both sides of the cylinder. 

\begin{figure}
	\centering    \includegraphics[width=0.8\textwidth]{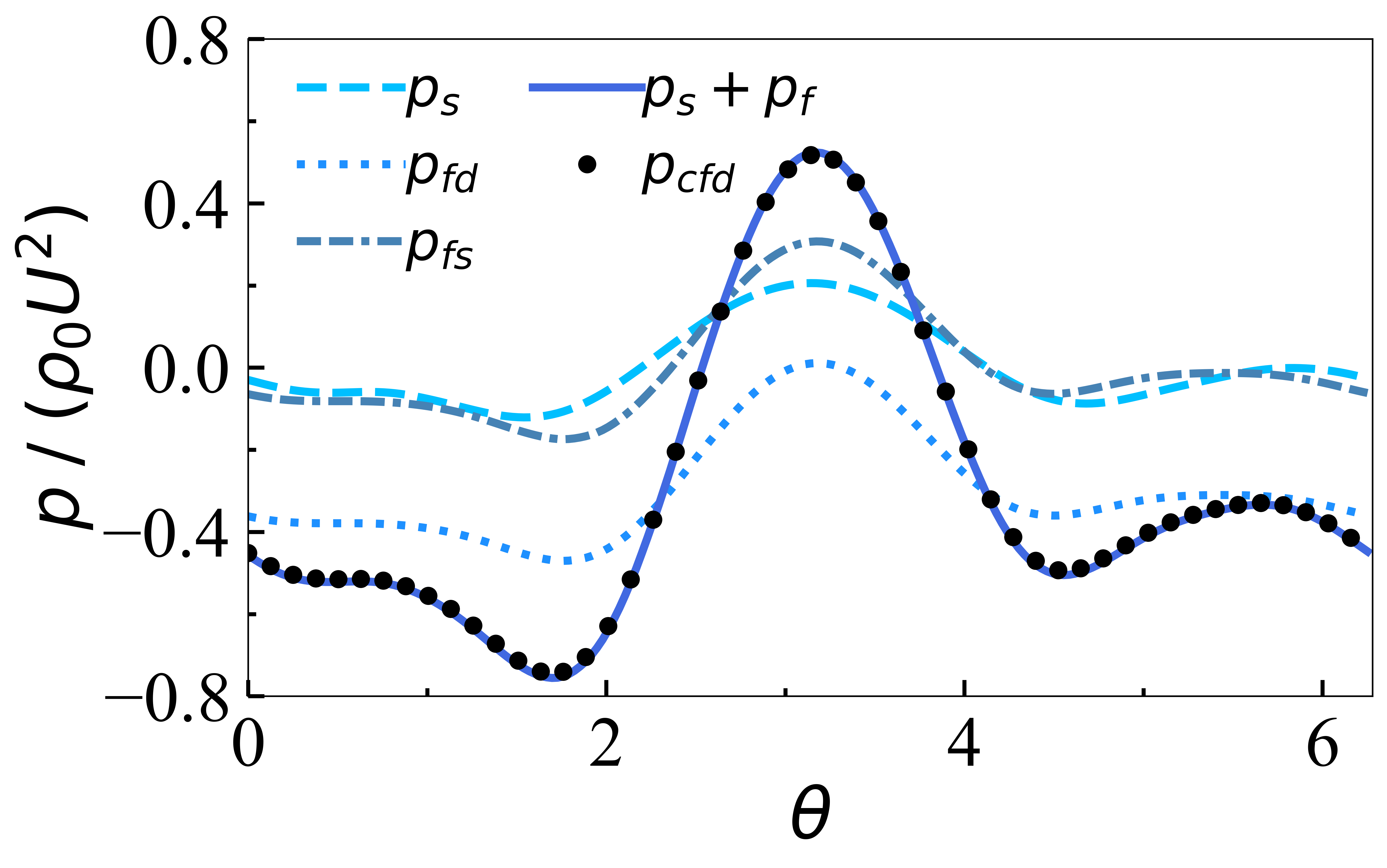}
	\caption{Comparison between the contribution from different sources to instantaneous pressure stress on the circular cylinder's surface}
	\label{fig12-2}
\end{figure}

Figure~\ref{fig12-2} shows the instantaneous pressure stress on the two-dimensional circular cylinder's surface correspond to the moment in Fig.~\ref{fig3} . As shown in the figure, the DFE method proposed in this work accurately estimates the pressure stress from convection volume sources and vorticity-acceleration surface pressure elements, demonstrating the ability of our method to handle two-dimensional flows.
The negative pressure on both sides of the rear stagnation point of the cylinder surface mainly originates from the direct radiation part of the convection volume source. The high pressure near the front stagnation point is primarily due to the scattering part from both the surface source and the convection volume source, with the latter's contribution being slightly larger. Additionally, the direct radiation part of the convection volume source at the front stagnation point is negative, which indicates that the positive contribution from direct radiation of volume sources near the front stagnation point is weaker than the negative contributions from the remaining volume sources.

\begin{figure}
	\centering    \includegraphics[width=1\textwidth]{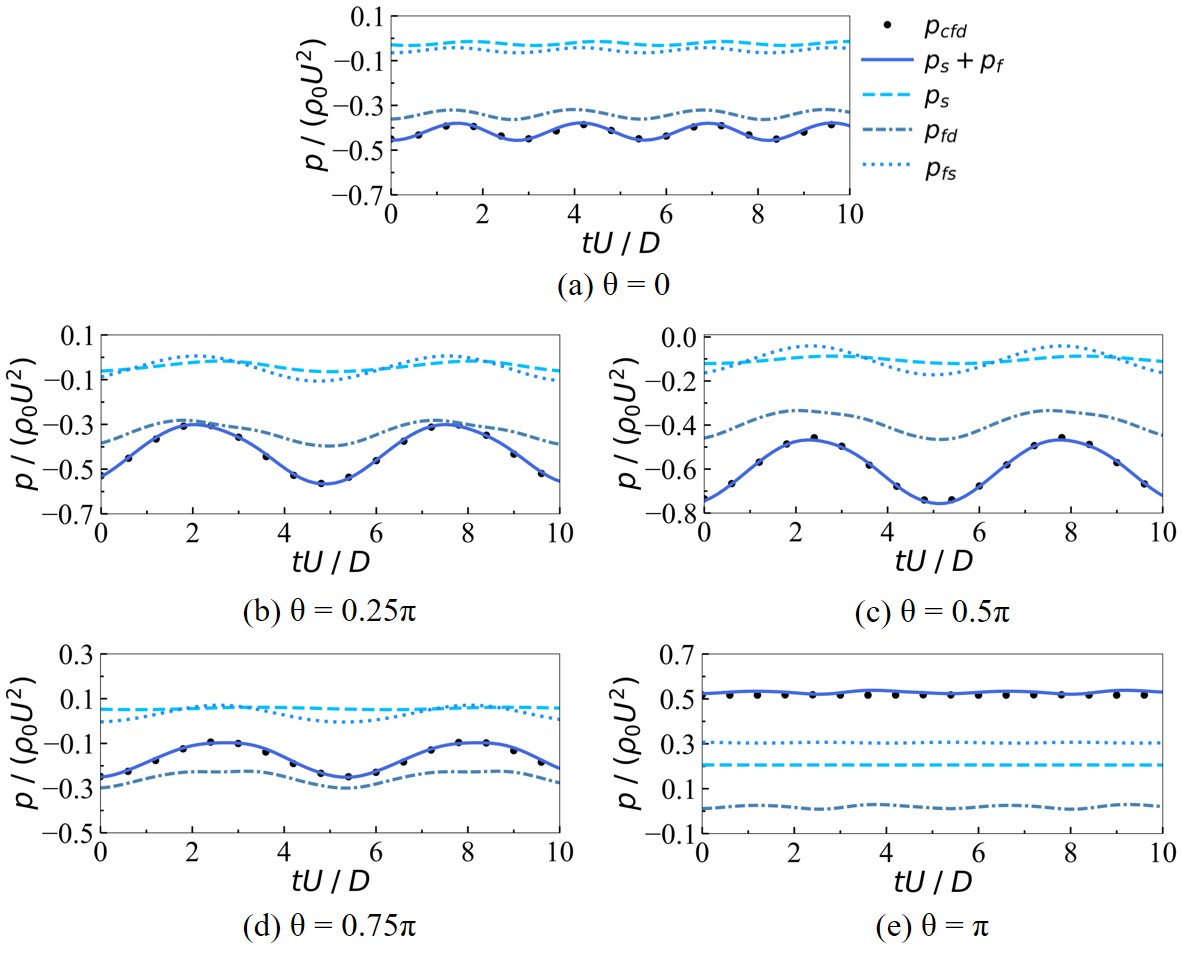}
	\caption{Comparison of the pressure computed by summing the contribution from convection volume source and vorticity-acceleration surface sources and standard results at (a) $\theta=0$ (b) $\theta=0.25\pi$ (c) $\theta=0.5\pi$ (d) $\theta=0.75\pi$ (e) $\theta=\pi$ on the solid boundary}
	\label{fig3-1}
\end{figure}

Figure~\ref{fig3-1} shows the composition of pressure at different locations on the surface of the circular cylinder. The solid dots denote directly sampled values from the simulated flow field, referred to as standard results hereinafter.
The solid lines represent the results calculated by Eq.~\eqref{eq8-1}, the dotted lines represent the contributions from the convection volume sources, and the dashed lines represent the contributions from the vorticity-acceleration surface sources. Here, $\theta=0$ indicates the rear stagnation point of the cylinder, and $\theta=\pi$ indicates the front stagnation point of the cylinder. 
From the $\theta=0$ to  $\theta=\pi$, the pressure fluctuation increases first and then decreases from the rear stagnation point to the front stagnation point.
This observation is consistent with that the fluctuation of drag is much smaller than that of lift.
The pressure fluctuation at different measurement points mainly comes from the convection volume source. The vorticity-acceleration surface source mainly contributes to the mean pressure. For instance, for the point at $\theta=0.5\pi$, the amplitude of the pressure fluctuation contributed by the surface source is approximately $\frac{1}{8}$ of that contributed by the volume source, while the ratio of mean pressure is about $\frac{1}{5}$. 
Furthermore, the direct radiation part of the convection volume source is closest to the pressure at each measurement point, except for the point at \(\theta = \pi\). This observation indicates that the direct radiation effect of the volume source is stronger than the wall scattering part. At the measurement point where \(\theta = \pi\), the contribution of the direct radiation part of the volume source term is close to zero, which suggests that the direct radiation parts of the positive and negative volume sources cancel each other out near the stagnation point.

%


\subsection{Two-dimensional flows over an oscillating circular cylinder}
\begin{figure}
	\centering    \includegraphics[width=1\textwidth]{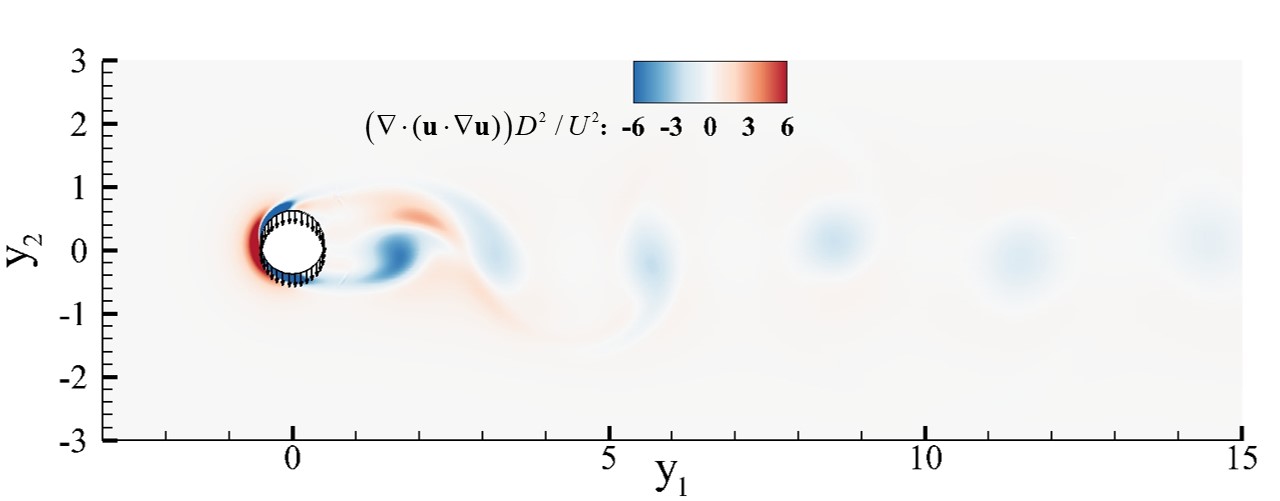}
	\caption{schematics of the flows over an oscillating circular cylinder and instantaneous volume source distribution}
	\label{fig7}
\end{figure}

To validate the proposed distributed force element method on moving boundaries, We simulate a two-dimensional flow over an oscillating circular cylinder, as shown in Fig.~\ref{fig7}. The Reynolds number based on the freestream speed and the diameter of the cylinder is ${\mathop{\rm Re}\nolimits}  = UD/\nu  = 185$, according to the work of~\cite{wang2025machine}. The circular cylinder is initialised at the origin and moves in the $y_2$ axis direction according to $y_c=0.2sin(0.98t)$ where $y_c$ is the time-varied location of circular cylinder's center.
More details about the setup can be found in our previous work~\citep{wang2025machine}. 
The virtual sources are positioned at $0.35D$ from the circular cylinder center. The integration domain of the volume source is defined as a rectangular region enclosing the cylinder. In this case, $y_{\rm left}$, $y_{\rm bottom}$ and  $y_{\rm top}$ is specified as $-2D$, $-4D$ and $4D$. 

\begin{figure}
	\centering    \includegraphics[width=1\textwidth]{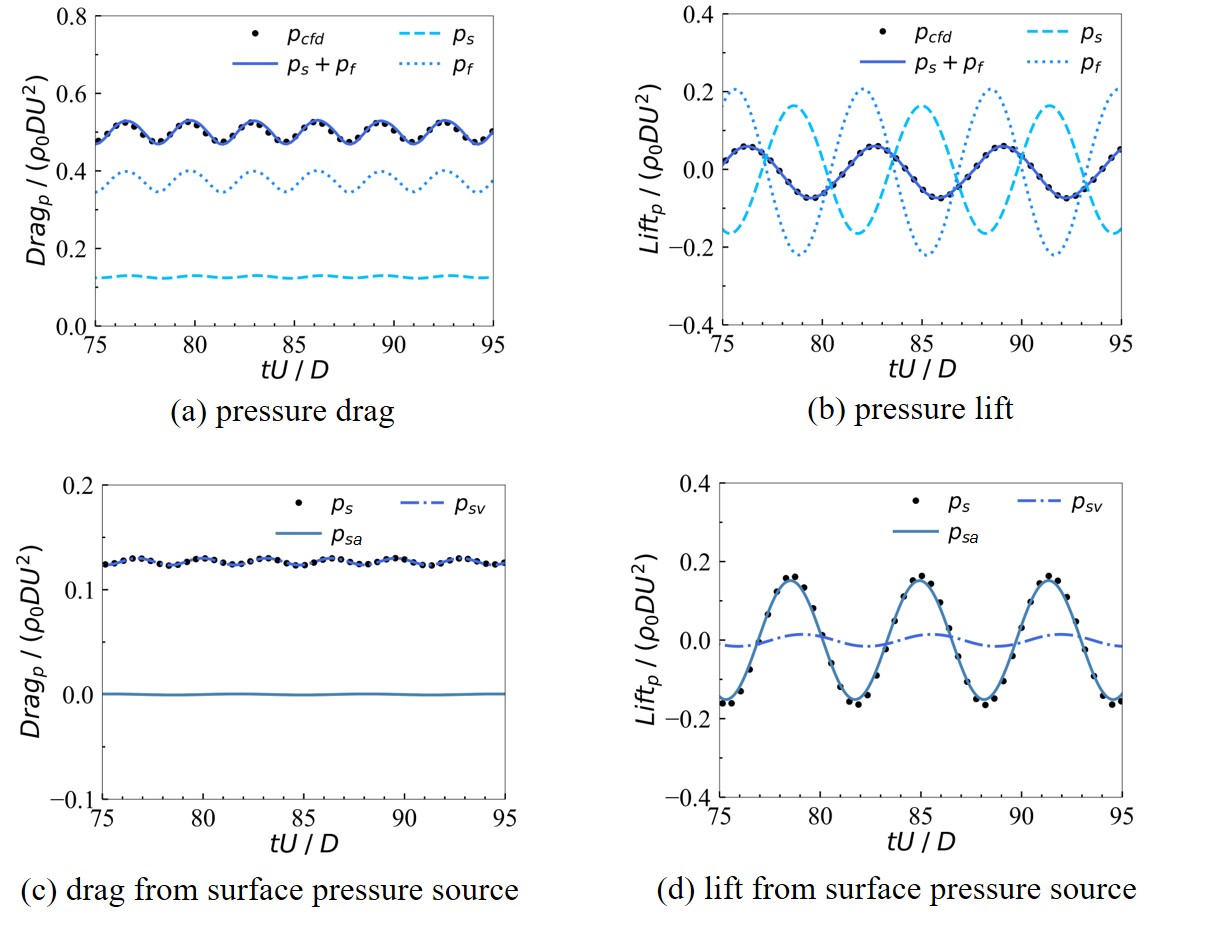}
	\caption{Comparison between the contribution from different sources to pressure drag and pressure by using the distributed force element method and directly solving NS equations: (a) contribution of different sources to  pressure drag; (b) contribution of different sources to  pressure lift; (c) contribution of different surface sources to pressure drag, and  (d) contribution of different surface sources to pressure lift;}
	\label{fig8}
\end{figure}

Figure~\ref{fig8} illustrates the temporal evolution of pressure drag and lift contributions from different pressure sources. The truncation position of the volume source is selected as $ y_{\text{right}} = 10D $. As shown in Figs.~\ref{fig8}a and~\ref{fig8}b, the results calculated by the DFE method align well with standard results. The pressure drag is dominated by volume source. For the pressure lift, the volume and surface sources partially cancel each other, resulting in a net lift smaller than either individual contribution.
 Fig.~\ref{fig8}c compares the contributions of acceleration and vorticity surface sources to pressure drag.  The acceleration sources contributes significantly less than the vorticity sources.  Fig.~\ref{fig8}d compares the contributions of acceleration and vorticity surface sources to pressure lift, the pressure lift from surface sources is dominated by the acceleration surface sources, suggesting that prescribing wall motion can modulate the amplitude of lift fluctuations.
 
\begin{figure}
	\centering    \includegraphics[width=1\textwidth]{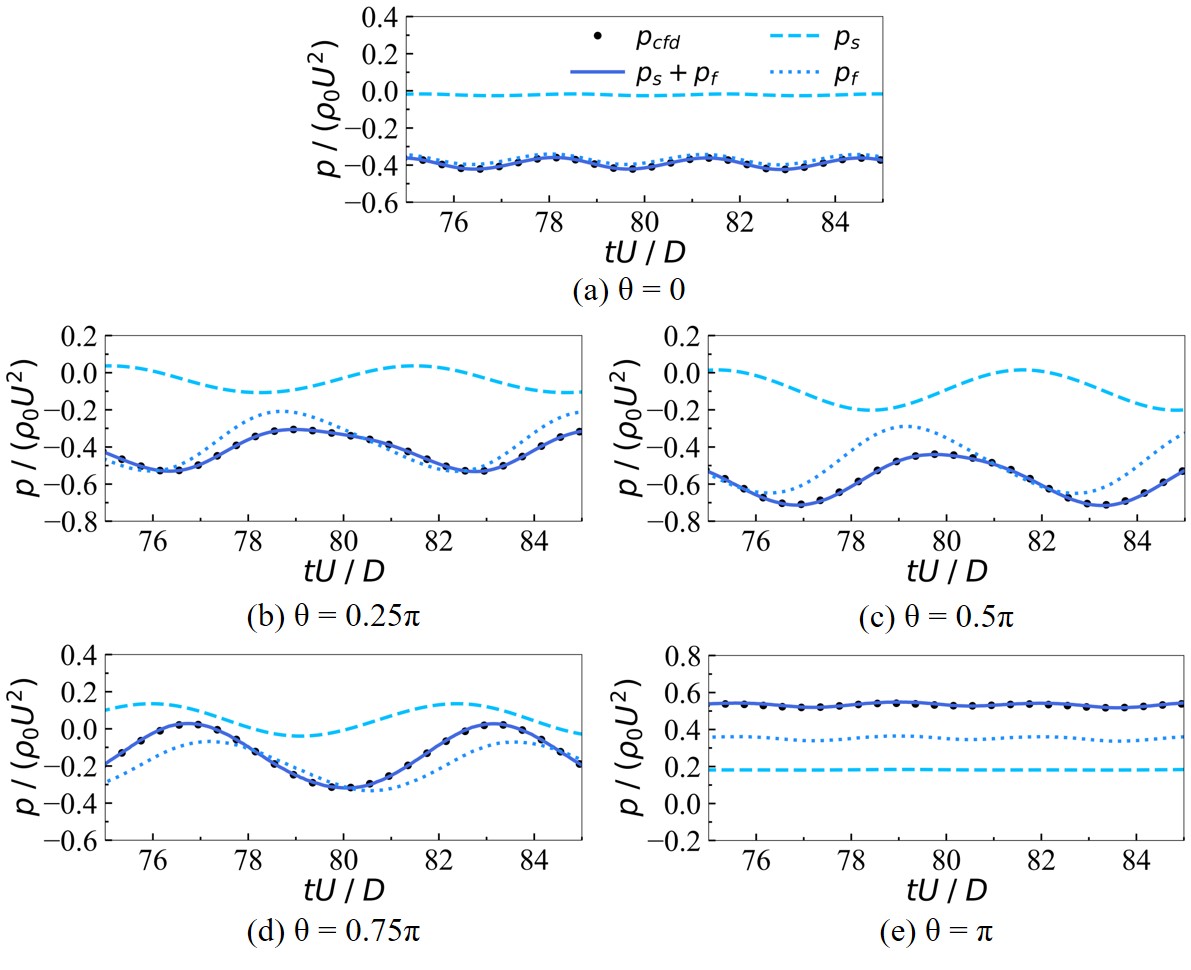}
	\caption{Comparison of the pressure computed by summing the contribution from convection volume source and vorticity-acceleration surface sources and standard results at  (a) $\theta=0$ (b) $\theta=0.25\pi$ (c) $\theta=0.5\pi$ (d) $\theta=0.75\pi$ (e) $\theta=\pi$ on the solid boundary}
	\label{fig8-1}
\end{figure}

Figure~\ref{fig8-1} shows the composition of pressure at different locations on the surface of the oscillating circular cylinder. The solid dots represent the results calculated by CFD, the solid lines represent the results calculated by Eq.~\eqref{eq8-1}, the dotted lines represent the contributions from the convection volume sources, and the dashed lines represent the contributions from the vorticity-acceleration surface sources.
It is obvious that due to the oscillation of the solid boundary, the vorticity-acceleration sources exhibit more intense fluctuations compared to the flow over a stationary circular cylinder.
The pressure fluctuation is small at the rear and front stagnation points on the solid boundary, similar to the flow over a stationary circular cylinder. However, the pressure fluctuation at point of $\theta=0.5\pi$ is no longer the strongest among the five points due to destructive cancellation between the pressure from surface and volume sources. In contrast, a constructive interference at $\theta=0.75\pi$ between the surface and volume sources can be observed, resulting in a enhancement of the pressure fluctuation at $\theta=0.75\pi$.

\begin{figure}
	\centering    \includegraphics[width=1\textwidth]{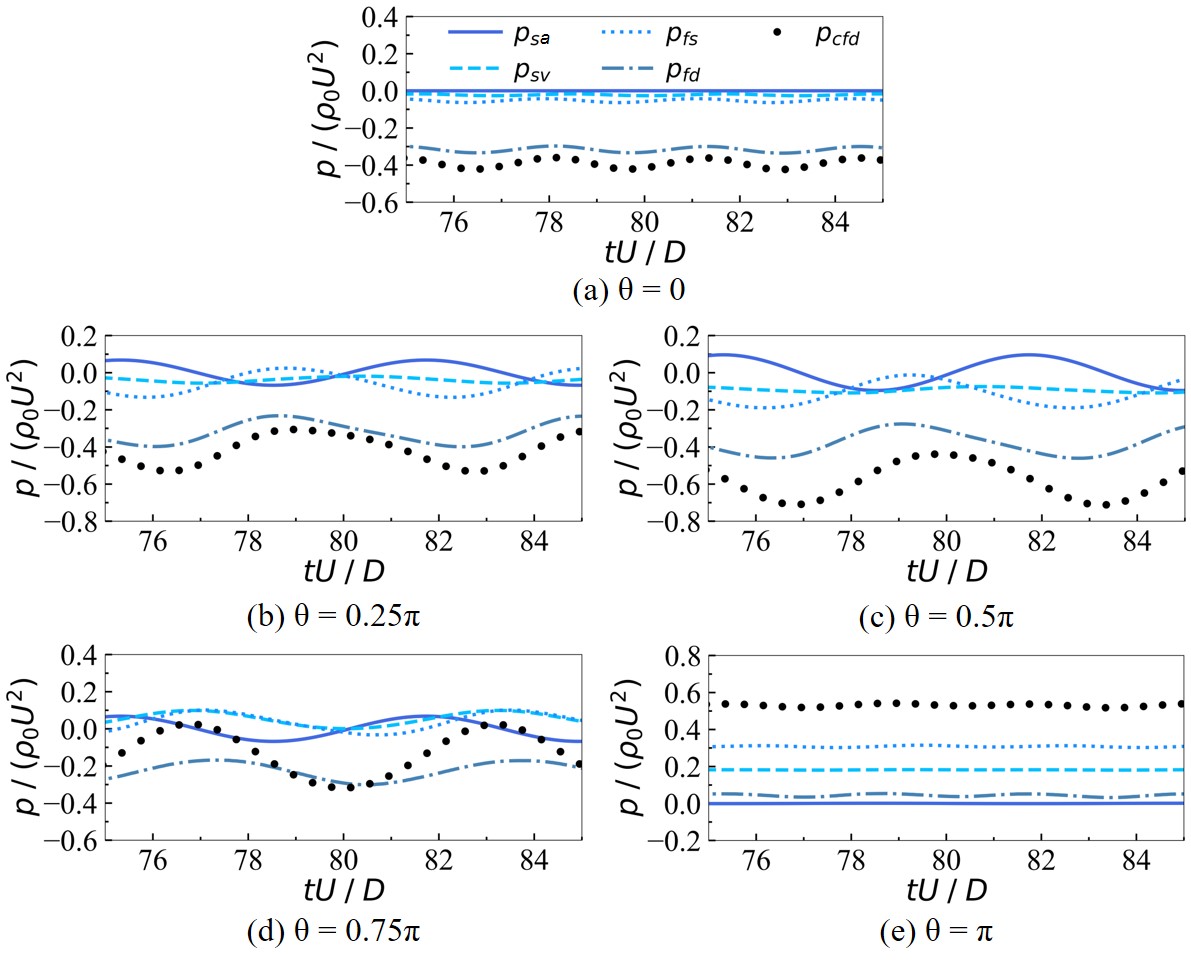}
	\caption{Comparison of the different components of pressure at  (a) $\theta=0$ (b) $\theta=0.25\pi$ (c) $\theta=0.5\pi$ (d) $\theta=0.75\pi$ (e) $\theta=\pi$ on the solid boundary}
	\label{fig8-2}
\end{figure}

Figure~\ref{fig8-2} shows the detailed composition of pressure at different locations on the surface of a cylinder. The solid dots represent the standard results,
the solid lines represent the contribution from acceleration surface sources, the dashed lines represent the contributions from the vorticity surface sources, the dotted lines represent the contributions from the scatter part of convection sources and the dash-dot line represent the contributions from the direct part of convection sources. 
At the point $\theta=0$,  the pressure is dominated by the direct part of the convection volume sources.
For the points at $\theta=0.25\pi$  and $\theta=0.5\pi$, the pressure fluctuation is significantly affected by the contribution from acceleration surface sources and direct and scatter part of the convection volume sources. Apparent destructive cancellation can be observed between the acceleration surface sources and direct and scatter part of the convection sources. 
For the point at $\theta=0.75\pi$, destructive cancellation happens between the contribution from acceleration and vorticity surface sources.
These observations suggest that the circular cylinder's motion effectively suppresses the pressure fluctuation along the vertical direction through the destructive cancellation between the contribution from acceleration surface sources with other sources. As a result, the pressure lift in this case is significantly  lower than that in the flow over a stationary circular cylinder.

\subsection{Laminar flows over a sphere}
To validate the proposed distributed force element method in three-dimensional conditions, We simulate a laminar flow over a sphere fixed in a uniform freestream. The diameter of the sphere is \(D\) and the freestream speed \(U\), respectively. The Reynolds number based on the freestream speed and the diameter of the cylinder is ${\mathop{\rm Re}\nolimits}  = UD/\nu  = 400$. 
The virtual sources are positioned at $0.3D$ from the sphere center. The integration domain of the volume source is defined as a rectangular region enclosing the sphere. In this case, the integration region of volume source along the $y_2$ and $y_3$ direction is specified as $-D$ to $D$. A pressure correction is used as the volume source region is truncated at $y_{\rm{right}}=4D$, where the details of pressure correction can be found in Appendix C.
\begin{figure}
	\centering    \includegraphics[width=1.0\textwidth]{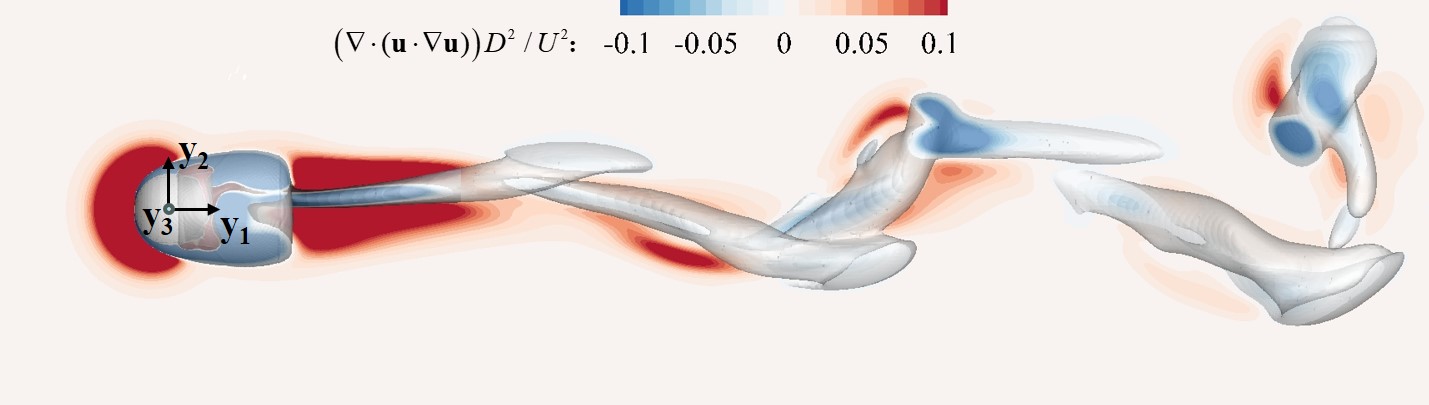}
	\caption{Schematics of the laminar flow over a sphere and instantaneous volume source distribution}
	\label{fig10}
\end{figure}

\begin{figure}
	\centering    \includegraphics[width=1\textwidth]{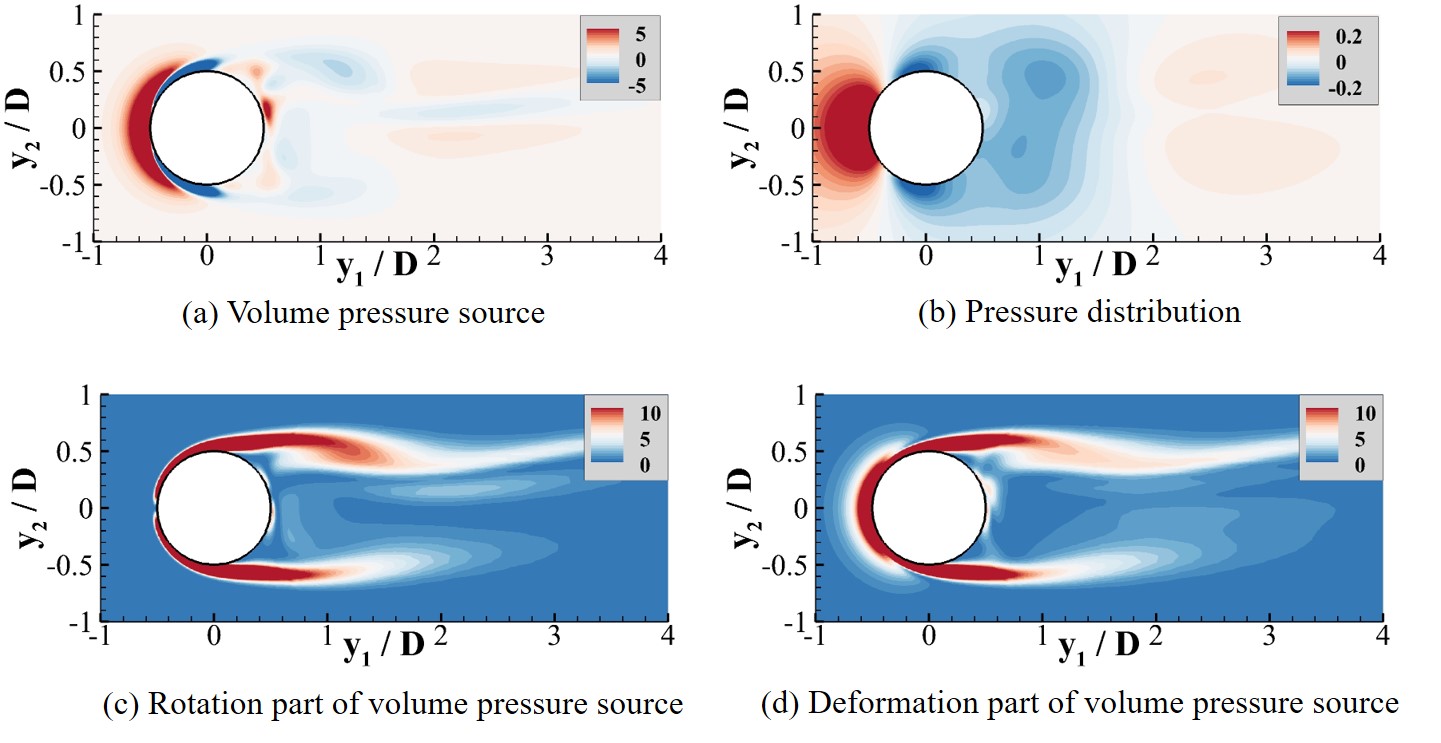}
	\caption{Comparison of instantaneous distribution of (a) volume source, (b) pressure, (c) rotation part of volume source, and (d) deformation part of volume source. }
	\label{fig11}
\end{figure}

Figure~\ref{fig10} shows the instantaneous flow field around a sphere. The vortex tubes are depicted by the isosurface of \( Q = 0.001 \), with the background colored according to the convection volume source $\left( {\nabla  \cdot ({\bf{u}} \cdot \nabla {\bf{u}})} \right)$. As shown in the figure, the volume source is primarily concentrated near the sphere and around the vortex tubes in the wake. 
The convection volume source is decomposed into two parts as $\left( {\nabla  \cdot ({\bf{u}} \cdot \nabla {\bf{u}})} \right) = {\left\| {\bf{E}} \right\|^2} - {\left\| {\bf{\Omega }} \right\|^2}$~\citep{wu2007vorticity} in Fig.~\ref{fig11}. ${{{\left\| {\bf{E}} \right\|}^2}}$ and ${{{\left\| {\bf{\Omega}} \right\|}^2}}$ are the squared Frobenius norm of deformation and rotation rate tensor of fluid elements, respectively. 
As shown in the Fig.~\ref{fig11}, there is a strong positive volume source near the stagnation point of the sphere, and strong negative volume sources on both sides of the sphere, which lead to the high-pressure region upstream the sphere and the low-pressure regions on the sides. However, the negative volume sources on both sides of the sphere decay rapidly after separation. As shown in Figs.~\ref{fig11}c and~\ref{fig11}d, the rotation and deformation in the shear layer of the sphere are relatively strong, but their intensity and coverage area are similar after the shear layer separation, which leads to the cancellation of the shear and deformation effects within the separated shear layer and the rapid decay of the volume source in the shear layer. Downstream of the sphere, the volume source is weaker, with negative volume sources dominating inside the vortex tubes and positive volume sources dominating on the sides of the vortex tubes.

\begin{figure}
	\centering    \includegraphics[width=1\textwidth]{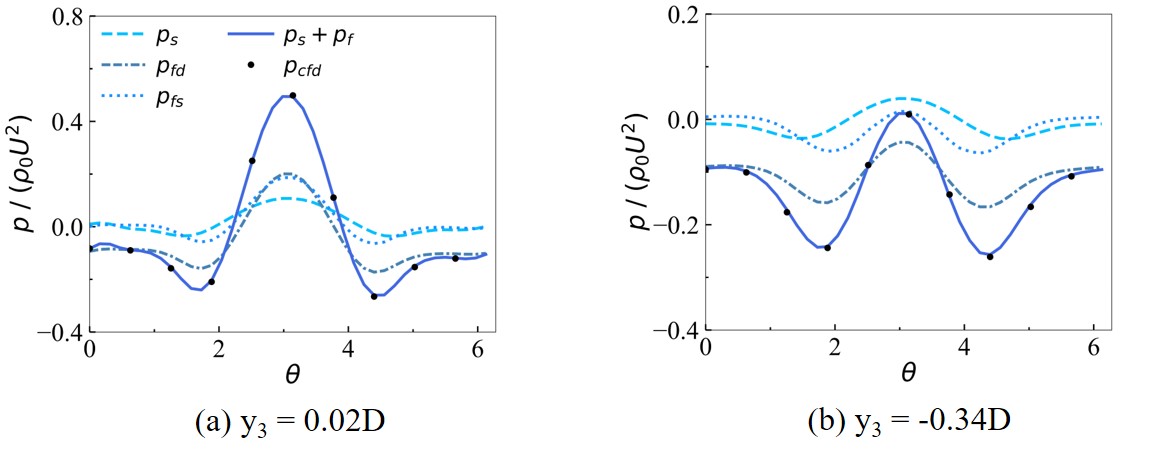}
	\caption{Comparison between the contribution from different sources to instantaneous pressure stress on the sphere at (a) $y_3=0.02D$ and (b) $y_3=-0.34D$;}
	\label{fig12-1}
\end{figure}

Figure~\ref{fig12-1} shows the instantaneous pressure distribution at $y_3=0.02D$ and $y_3=-0.34D$ on the sphere surface. As shown in the figure, the DFE method proposed in this work accurately estimates the pressure stress from convection volume sources and vorticity-acceleration surface pressure elements, demonstrating the ability of our method to handle three-dimensional flows. The volume source is the main contributor at $y_3=0.02$ and $y_3=-0.34D$. 
Compared to $y_3 = 0.02D$, the pressure at the front stagnation point is significantly reduced at $y_3 = -0.34D$. This reduction is primarily due to the decreased contribution from the convection volume source. This observation suggests that as $y_3$ changes from $0.02D$ to $-0.34D$, the relative contribution of the negative volume source within the boundary layer gradually increases, while the relative contribution of positive volume source close to the stagnation point gradually decreases. 

\subsection{Turbulent flows over a sphere}
We simulate a three-dimensional critical flow over a sphere fixed in a uniform freestream to validate the proposed distributed force element method for turbulent flows. The Reynolds number based on the freestream speed and the diameter of the cylinder is ${\mathop{\rm Re}\nolimits}  = UD/\nu  = 3700$. 
The virtual sources are positioned at $0.3D$ from the sphere center. The integration domain of the volume source is defined as a rectangular region enclosing the sphere. In this case, the integration region of volume source along the $y_2$ and $y_3$ direction is specified as $-D$ to $D$. 
\begin{figure}
	\centering    \includegraphics[width=1.0\textwidth]{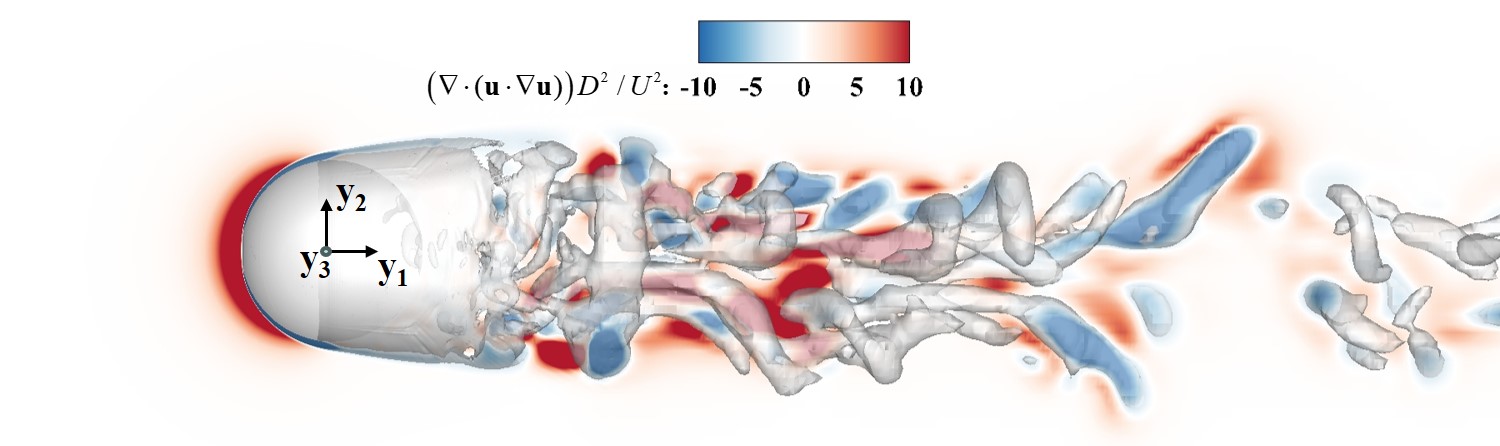}
	\caption{Schematics of the turbulent flow over a sphere and instantaneous volume source distribution}
	\label{fig14}
\end{figure}

\begin{figure}
	\centering    \includegraphics[width=1.0\textwidth]{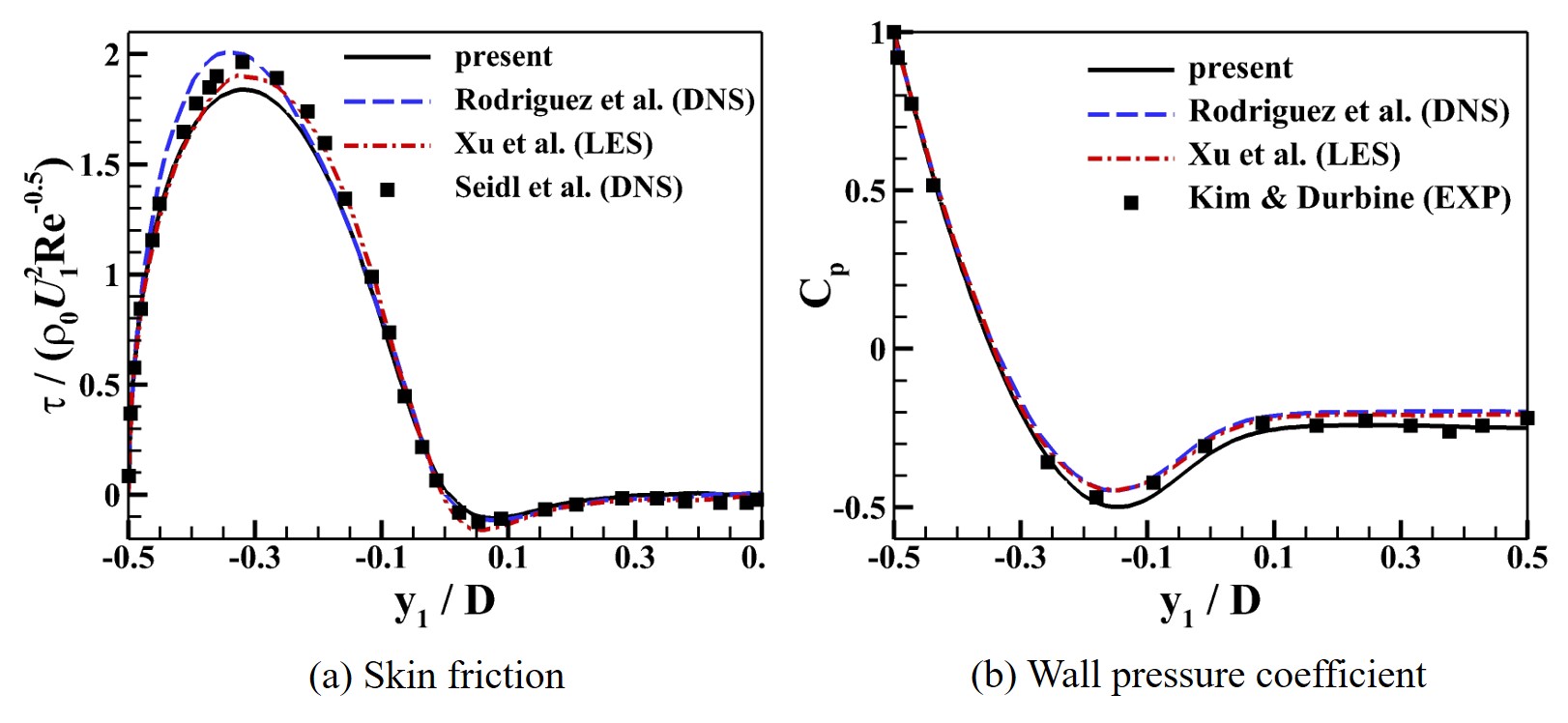}
	\caption{Mean profiles around the sphere surface: (a) pressure coefficient and (b) skin friction.}
	\label{fig15}
\end{figure}

Figure~\ref{fig14} shows the instantaneous turbulent flow around the sphere. The vortex tubes are depicted by the isosurface of \( Q = 1 \), with the background colored according to the volume source \( \left\| \mathbf{E} \right\|^2 - \left\| \mathbf{\Omega} \right\|^2 \). Similar to the flow at $Re= 400$, the volume source is primarily concentrated near the sphere and around the vortex tubes in the wake.  The mean pressure coefficient and skin friction at the cross-section $y_3=0$ matches well with different numerical and experimental results~\citep{rodriguez2011direct,xu2018novel,kim1988observations,seidl1997parallel}, as shown in Fig.~\ref{fig15}.

\begin{figure}
	\centering    \includegraphics[width=1.0\textwidth]{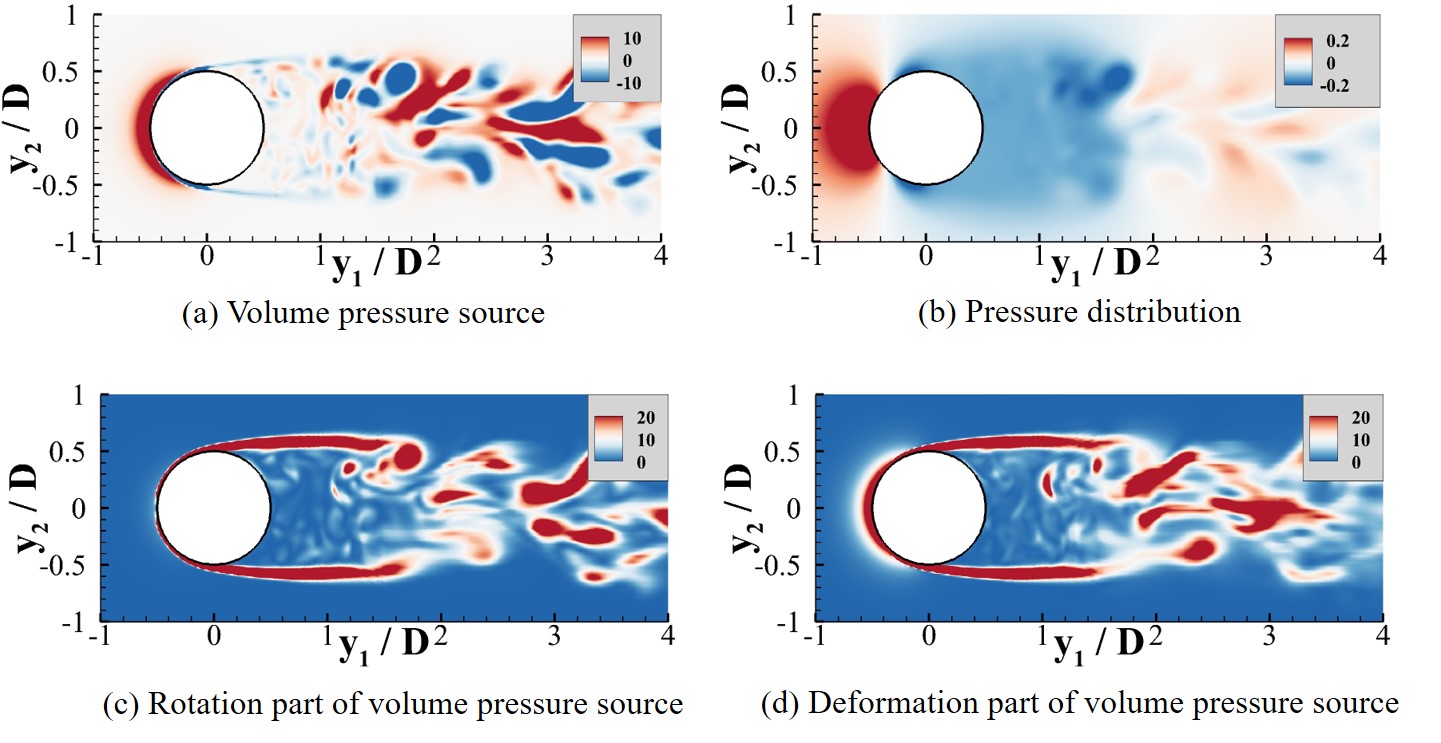}
	\caption{Comparison of instantaneous distribution of (a) volume source, (b) pressure, (c) rotation part of volume source, and (d) deformation part of volume source.}
	\label{fig16}
\end{figure}

Figure~\ref{fig16} further provides the distributions of the volume source, pressure, and components of the volume source. As shown in the figure, the negative volume sources in the shear layer decay rapidly after separation, which is similar to the flow at $Re= 400$.  Cancellation of the rotation and deformation volume sources within the separated shear layer is observed after the shear layer separation, leading to the rapid decay of the volume source in the shear layer.   The volume source is strong downstream the shear layer due to interaction between vortex tubes, resulting in strong local pressure in the wake. 
Between the separated shear layer and the downstream vortexes,  a region with small volume source is observed. As a result, the pressure at the rear of the sphere mainly comes from the strong volume sources at a certain distance from the rear. This leads to a relatively uniform pressure near the rear.

\begin{figure}
	\centering    \includegraphics[width=1.0\textwidth]{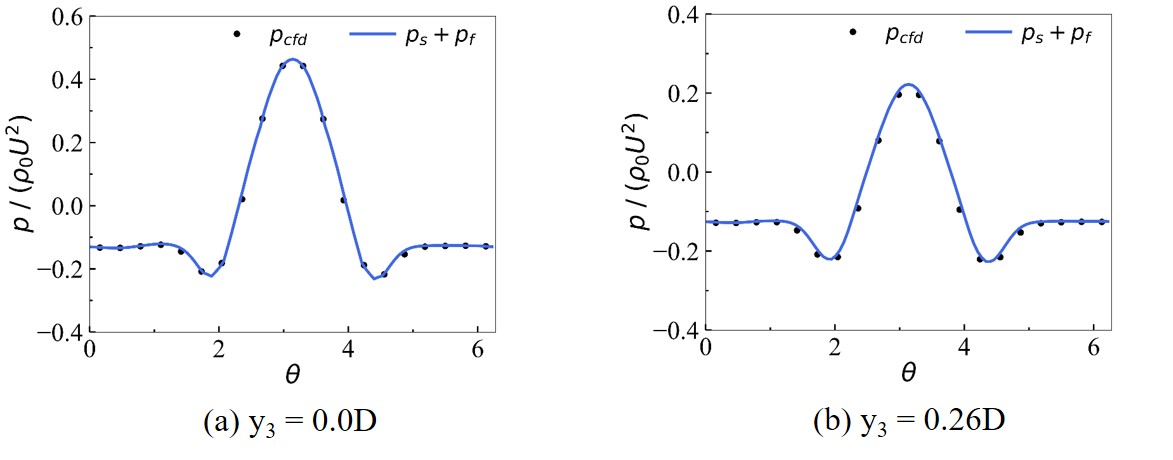}
	\caption{Comparisons between the instantaneous pressure distribution at different $y_3$ cross-section on the sphere computed using the DFE method with standard results at
 (a) $y_3=0.0D$ ; (b) $y_3=0.26D$.}
	\label{fig17}
\end{figure}
\begin{figure}
	\centering    \includegraphics[width=1.0\textwidth]{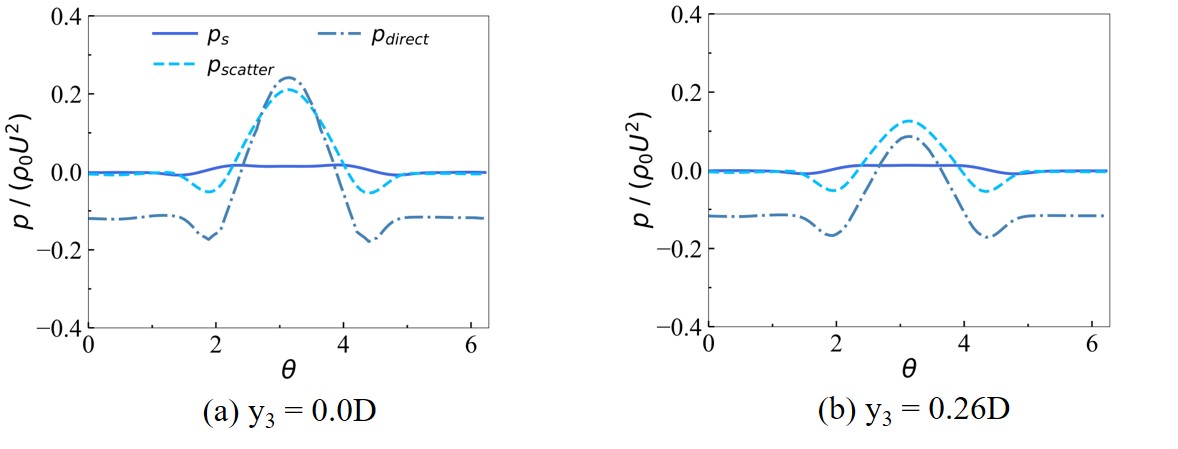}
	\caption{Comparison between the contribution from different sources to instantaneous pressure stress at (a) $y_3=0.0D$ and (b) $y_3=0.26D$;}
	\label{fig18}
\end{figure}

Figure~\ref{fig17} shows the instantaneous pressure distribution at $y_3=0$ and $y_3=0.26D$ on the sphere surface. As shown in the figure, the method proposed in this work accurately estimates the pressure stress from volume sources and surface pressure elements, demonstrating the ability of our method to handle turbulent flows. Figure~\ref{fig18} illustrates the different components of pressure stress. As shown in the figure, the volume source is the main contributor at $y_3=0.0$ and $y_3=0.26D$. The negative pressure at the rear of the sphere mainly comes from the direct radiation part, and the contributions of the direct radiation part and the wall scattering part to the positive pressure at the front of the sphere are close. This observation suggests that, the wall scattering part is more sensitive to the volume sources near the wall compared with the direct radiation part, confirming the rationality of the pressure correction in appendix.

\section{Conclusion}

We have developed a distributed force element method based on fundamental solutions approach to decompose the force on a body moving in an incompressible fluid. 
The distribution of force at any point on the wall is considered as the superposition of contributions from volume sources and surface elements. The volume sources accounts for the convection of fluid elements, which can also be decomposed to account for the deformation and rotation of fluid elements. The surface integrals, on the other hand, account for the acceleration of solid boundaries and the influence of boundary vorticity. Based on the idea of Green's functions, we introduce the method of fundamental solutions to determine the contributions of different surface elements and volume sources to the pressure at any point on the wall. The volume source term can be further decomposed into a direct radiation part and a wall scattered part by decomposition the Green's function.

We first utilize the two-dimensional flow around a circular cylinder at a Reynolds number of 150 to test the applicability of the proposed distributed force element method.
In the low Reynolds number flow around a two-dimensional cylinder, we find that fluctuation of pressure mainly comes from the convection volume sources.
We then validate the proposed method for moving objects and three-dimensional cases using the flow around an oscillating cylinder and the flows around a three-dimensional sphere. 
The results show that the pressure computed by using the proposed distributed force element method matches well with standard results. In the flow over an oscillating circular cylinder, we find that the pressure drag and its fluctuation mainly come from the contribution of volume sources. 
For the points on the surface, destructive and constructive interferences between the contribution from acceleration surface sources and other sources lead to the deviation of amplitude of pressure fluctuation from the flow over a stationary circular cylinder. 
As a result, the pressure lift is significantly lower than that in the flow over a stationary circular cylinder due to the destructive interferences.
This implies that we can use phase cancellation to control pressure fluctuation by adjusting the form of the cylinder's vertical oscillation. 
In the three-dimensional sphere flow, we find that volume sources are mainly concentrated near the stagnation point of the sphere, inside the wake vortex, and on its sides. For the shear layers on both sides of the sphere, only the attached parts have significant negative volume sources, leading to local low-pressure areas. Although the detached shear layers have strong fluid element rotation and deformation, these two effects almost cancel each other out, resulting in weak volume sources in the detached shear layers.

\section*{Acknowledgements}
This work is supported by the NSFC Excellence Research Group Program for ‘Multiscale Problems in Nonlinear Mechanics’ (No. 12588201), the National Natural Science Foundation of China (Nos. 12425207, 12425207 and 92252203), the Chinese Academy of Sciences Project for Young Scientists in Basic Research (Grant No. YSBR-087), and the Strategic Priority Research Program of Chinese Academy of Sciences (Grant No. XDB0620102). The computations are conducted on Tianhe-1 at the National Supercomputer Center in Tianjin.

 \section*{Declaration of interests}
 The authors declare that they have no known competing financial interests or personal relationships that could have appeared to influence the work reported in this paper.

\section*{Appendix A: Validation for the laminar flow over a circular cylinder}
This appendix uses flow over a two-dimensional circular cylinder to validate the proposed distributed force element (DFE) method. The resultant force computed from DFE are compared with standard results. The convergence for the size of the convection volume source's integration region is examined.

The resultant force vector $\mathbf{F}$ is defined as the sum of pressure difference and friction contributions:
\begin{equation}\label{eqa1}
\begin{split}
F_i = \left( \int_S p_s n_i \, dS + \int_V p_f n_i \, dV \right) + \int_S \tau n_i \, dS,
\end{split} \tag{A1}
\end{equation}
where $\tau$ denotes the wall shear stress. $F_i$ represents the resultant force along the $y_i$-direction.
Substituting into Eq.~(\ref{eq8-1}) and expressing wall shear stress in terms of boundary vorticity, the resultant force expands to:
\begin{equation}\label{eqa2}
\begin{split}
F_i =& \int_S {\left( {\int_V { - {\left( {\nabla  \cdot ({\bf{u}} \cdot \nabla {\bf{u}})} \right)}} {G_f}{\mkern 1mu} {\kern 1pt} dV} \right)} {n_i}{\mkern 1mu} {\kern 1pt} dS \\
&+ \int_S \left( -\int_S \frac{1}{\mathrm{Re}} (\nabla \times \boldsymbol{\omega}) \cdot \mathbf{n} \, G_s \, dS \right) n_i \, dS \\
&+ \int_S \left( -\int_S \left(  \frac{D\mathbf{u}}{Dt} \cdot \mathbf{n} \right) G_s \, dS \right) n_i \, dS \\
&+ \int_S \frac{1}{\mathrm{Re}} (\mathbf{n} \times \boldsymbol{\omega})_i \, dS.
\end{split}\tag{A2}
\end{equation}
We define volume integral and surface integral terms as
$\mathbf{F}_f \equiv \int_V p_f n_i \, dV$  and $\mathbf{F}_s \equiv \int_S p_s n_i \, dS + \int_S \tau n_i \, dS$, respectively.

\begin{figure}
	\centering    \includegraphics[width=1\textwidth]{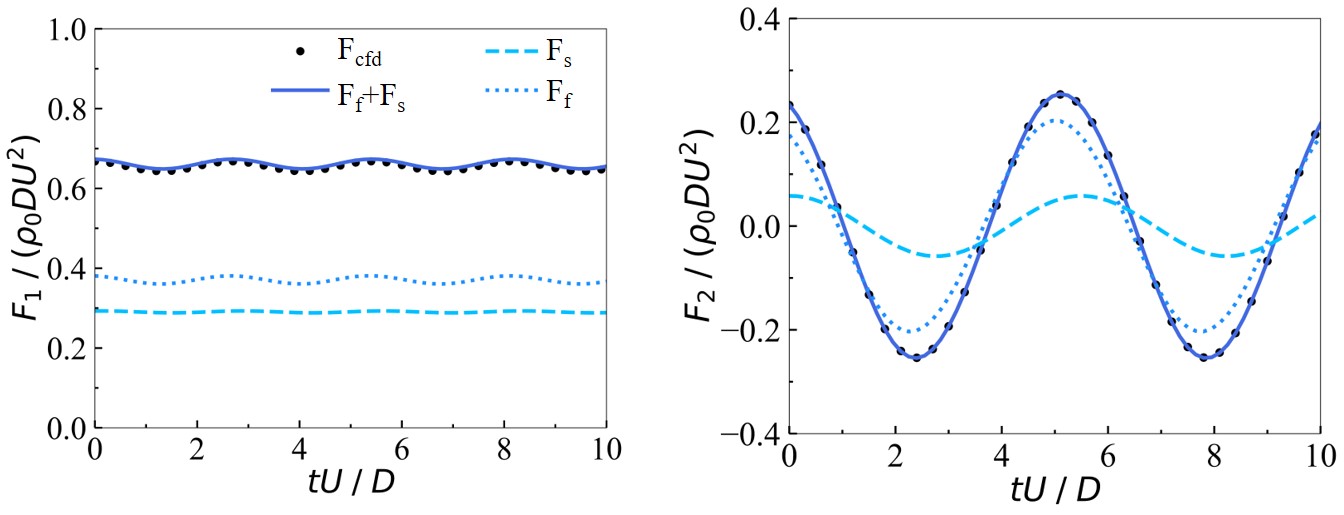}
	\caption{Comparison Drag (a) and Lift (b) computed by using the distributed force element method and directly solving NS equations }
	\label{fig2}
\end{figure}

Figure~\ref{fig2} shows the total drag and lift calculated using the distributed force element method. As shown in the figure, the total drag and lift calculated by Eq.~(\ref{eqa2}) are in good agreement with standard results. The lift fluctuation is much larger than that of the drag, and mainly comes from the volume source. 

\begin{figure}
	\centering    \includegraphics[width=1.0\textwidth]{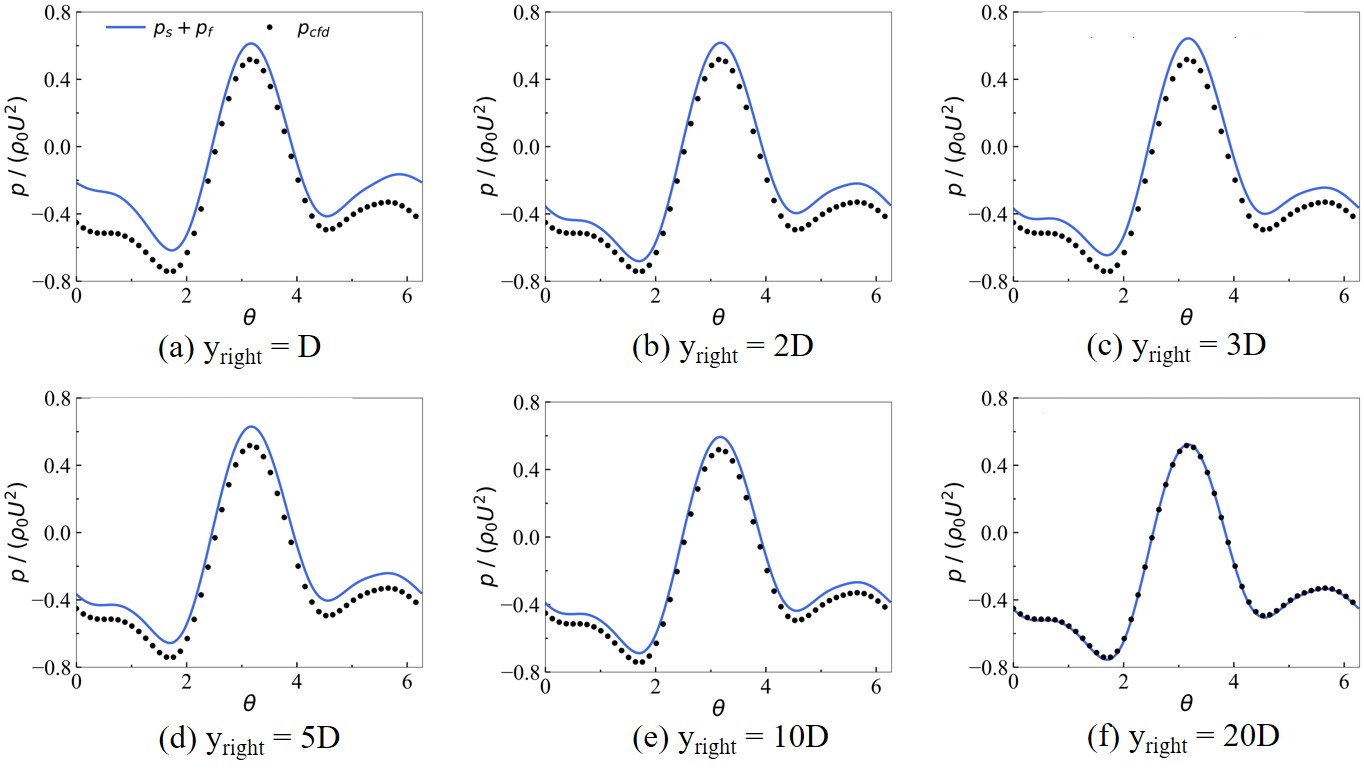}
	\caption{Comparison of instantaneous pressure stress distributions computed with different streamwise truncation positions.} 
	\label{fig4}
\end{figure}

Figure~\ref{fig4} compares the effects of different truncation positions $y_{\text{right}}$ of the integration domain on the computed pressure stress on the solid boundary. As the truncation position moves gradually downstream, the pressure stress distribution obtained by Eq.~(\ref{eq8-1}) gradually approaches the standard results. 

\section*{Appendix B: Pressure correction for the finite integration region}
\begin{figure}
	\centering    \includegraphics[width=1.0\textwidth]{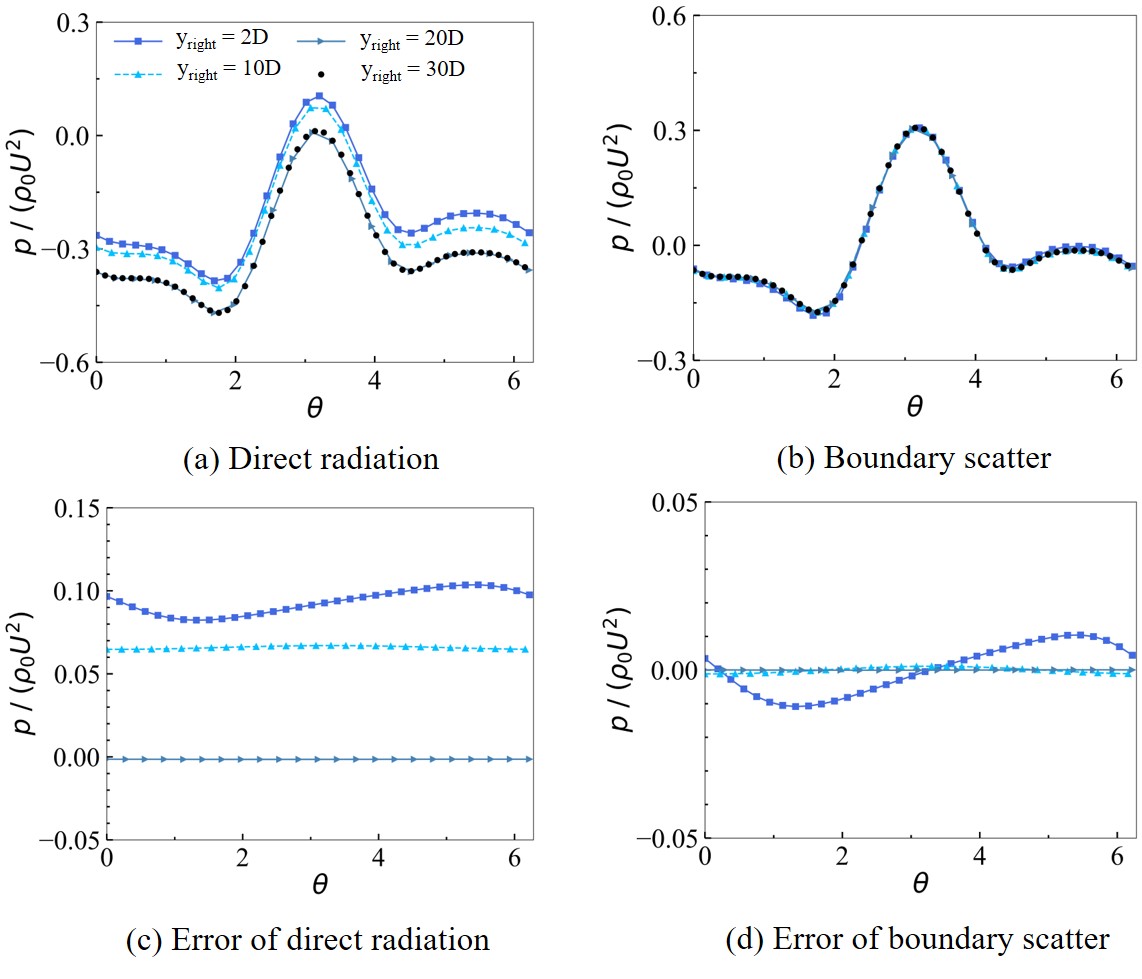}
	\caption{Contribution from direct radiation and boundary scatter from the volume source to the instantaneous pressure stress distribution and related absolute error: (a) contribution from direct radiation; (b) contribution from boundary scatter; (c) absolute error of the direct radiation; (d) absolute error of the boundary scatter.}
	\label{fig5}
\end{figure}
This appendix proposes a distributed force correction based on single-point pressure stress measurements to further reduce the computational cost of the distributed force element method. The cases employed in this appendix are identical to that utilized in Appendix B.

The source of the truncation error is identified by separating the direct radiation part and the wall scattering part of the volume source term, as shown in Fig.~\ref{fig5}.
The direct radiation part has obvious errors when the integration domain is not large enough compared with the wall scattering part. 
As can be seen in Figs.~\ref{fig5}c and~\ref{fig5}d, the errors of direct radiation part does not vary significantly with the angle. The contribution of the direct radiation part to the pressure at any point on the wall can be written as
\begin{equation}\label{eqb1}
\begin{split}
{p_f} = \int_{{V_R}} {-{\left( {\nabla  \cdot ({\bf{u}} \cdot \nabla {\bf{u}})} \right)}} {G}dV.
\end{split}\tag{B1}
\end{equation}
In two-dimensional cases, the Green's function for Poisson eqution is $G = \frac{1}{{2\pi }}\ln \left( {r} \right)$, where $r$ denotes the distance from the volume source to the wall. When the volume source is sufficiently far away from the wall, i.e., $r \gg D$, the distance from the volume source to any point on the wall is approximately equal to the distance to the center of the cylinder. Therefore, the direct radiation part of the pressure contribution from volume sources far from the wall to any point on the wall is nearly the same. As a result, the error caused by the truncation of the volume source does not vary significantly with the angular region. Based on this, we can make a correction for the pressure at a point on the wall when the integration domain is finite, that is
\begin{equation}\label{eqb2}
\begin{split}
p = \int_{{V_R}} { -{\left( {\nabla  \cdot ({\bf{u}} \cdot \nabla {\bf{u}})} \right)}} {G_f}dV - \int_S {\frac{1}{{{\rm{Re}}}}\left( {\nabla  \times {\bf{\omega }}} \right)}  \cdot {\bf{n}}{G_s}dS - \int_S {\left( {\frac{{D{\bf{u}}}}{{Dt}} \cdot {\bf{n}}} \right)} {G_s}dS + {p_{{\rm{correct}}}},
\end{split}\tag{B2}
\end{equation}
where ${p_{\text{correct}}}$ is obtained by matching the pressure at a point on the wall.

\begin{figure}
	\centering    \includegraphics[width=1\textwidth]{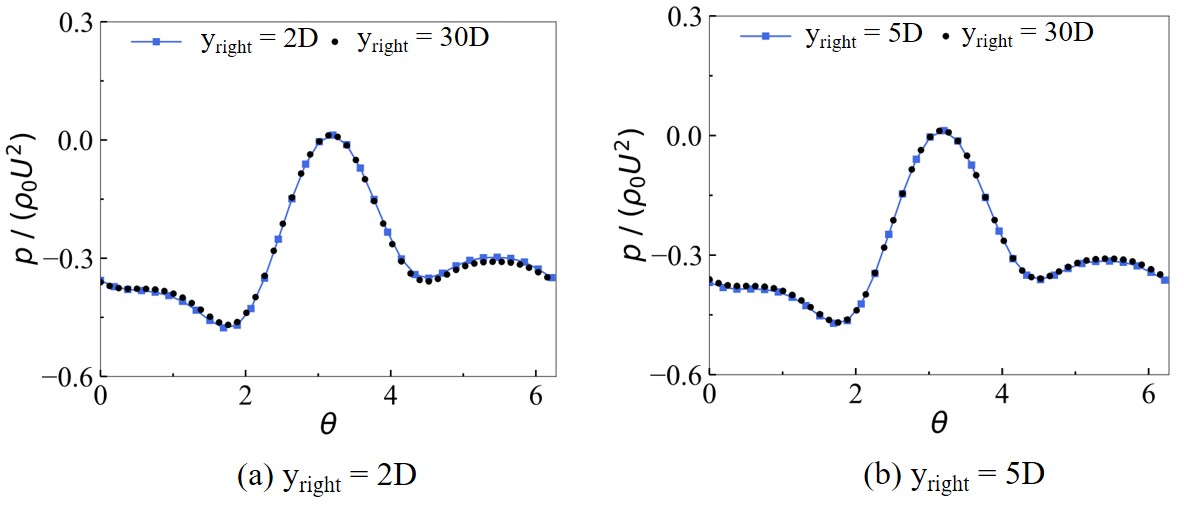}
	\caption{Corrected pressure stress distribution by using ~\eqref{eqb2} with different streamwise truncation position: (a) $y_{right}=2D$ and (b) $y_{right}=5D$.}
	\label{fig6}
\end{figure}

Figure~\ref{fig6} shows the results of the instantaneous pressure distribution calculated using the correction formula Eq.~(\eqref{eqb2}), where ${p_{\text{correct}}}$ is calculated by matching the pressure at the forward stagnation point of the circular cylinder. As shown in the figure, the corrected instantaneous pressure distribution using formula Eq.~(\eqref{eqb2}) agrees well with the results obtained when the integration domain is sufficiently large. This indicates that the contribution of the volume sources outside the truncated integration domain can be corrected by matching the pressure stress on the solid boundary.
\clearpage
\bibliographystyle{jfm}
\bibliography{jfm}

\end{document}